\documentclass[12 pt,aps,prd,showpacs]{revtex4}   
\usepackage{amsmath}    
\usepackage{graphicx}   
\begin{document}
\title{Time-dependent Robin boundary conditions in the dynamical Casimir effect}
\author{C Farina$^{1}$, Hector O. Silva$^{2}$, Andreson L. C. Rego$^{1}$ and Danilo T. Alves$^{2}$}
\affiliation{
(1) - Instituto de F\'\i sica, Universidade Federal do Rio de Janeiro, Caixa Postal 68528, 21945-970, Rio de Janeiro, RJ, Brazil\\
(2) - Faculdade de F\'\i sica, Universidade Federal do Par\'a, 66075-110, Bel\'em, PA,  Brazil}
\date{\today}
\begin{abstract}
Motivated by experiments in which moving boundaries are simulated by time-dependent properties of static systems, we discuss the model of a massless scalar field submitted to a time-dependent Robin boundary condition (BC) at a static mirror in $1+1$ dimensions. Using a perturbative approach, we compute the spectral distribution of the created particles and the total particle creation rate, considering a thermal state as the initial field state.
\end{abstract}
\pacs{03.70.+k, 11.10.Wx, 42.50.Lc}
\maketitle

\section{Introduction}
\label{introducao}

The conversion of virtual particles into real ones by moving boundaries  in vacuum is known as dynamical Casimir effect (DCE). The radiation reaction force acting on a neutral moving plate is another aspect of this quantum vacuum effect. 
The first theoretical prediction of the DCE was made by Moore approximately $40$ years ago \cite{Moore}, in the context of a scalar field in $1+1$ dimensions, in a cavity and subjected to Dirichlet boundary condition (BC).
Other pioneering papers on the DCE were done by DeWitt \cite{DeWitt} and by Fulling and Davies \cite{Fulling-Davies}.

Analogously to what happens in ordinary quantum mechanics, where a system initially in the fundamental state can jump into an excited state due to the interaction with an external time-dependent potential, a quantized field can also leave the vacuum state and jump into an excited state due to the interaction with an external time-dependent potential.
In the DCE, moving boundaries can be considered as external time-dependent potentials (for example, a moving boundary can be described by an electric permittivity or a magnetic permeability changing in time). For this reason, the interaction between quantized fields and moving mirrors induces the field to go out of the vacuum state. In other words, moving boundaries may give rise to particle creation. By arguments of energy conservation, we expect that the energy of the created particles is
taken from the mechanical energy of the moving boundary. Similar to what occurs in classical electromagnetism - where the radiation emitted by an accelerated charge  is always accompanied by a radiation reaction force on that charge - here we also have a radiation reaction force on a moving boundary due to particle creation.

Although the Casimir force on a single plate at rest is zero, the corresponding fluctuations of this static force do not vanish \cite{Barton-1991}. This fact suggests a possible way of understanding the DCE  in terms of the fluctuation-dissipation theorem \cite{flutuacao-dissipacao}. In the linear response theory, an appropriate application of this theorem to the problem of a moving mirror in vacuum leads to a dissipative force on the plate, which is, in turn, proportional to the fluctuations of the static force \cite{Braginsky-Khalili-1991,Jaekel-Reynaud-1992}. Other works also discussed the DCE via this point of view (see Refs. \cite{MaiaNeto-Reynaud-1993,Dalvit-Maia-Neto-Mazzitelli-arXiv-1006-4790v1-2010}).
 Again, invoking the energy conservation, the dissipative forces on moving boundaries is related to the conversion of mechanical energy into field energy (real particle creation).

Since the pioneering papers of Moore, DeWitt, Fulling and Davies, several authors have studied the DCE in different contexts and using different procedures (see Refs. \cite{Dalvit-Maia-Neto-Mazzitelli-arXiv-1006-4790v1-2010,V-V-Dodonov-Review} for a nice review):
perturbative \cite{Ford-Vilenkin-PRD-1982} and exact \cite{Cole-Schieve-1995} approaches;
distinct fields and BCs 
\cite{Alves-Farina-Maia-Neto-JPA-2003,Mintz-Farina-Maia-Neto-Robson-JPA-2003-I,Mintz-Farina-Maia-Neto-Robson-JPA-2003-II,Maia-Neto-LAS-Machado-PRA-1996};
single mirrors \cite{single-mirrors} and cavities \cite{cavities}.

Motivated by the first experimental observation of the DCE, made by Wilson and collaborators \cite{Wilson-Johansson-et-al-arXiv-1105.4720-2011} in the context of circuit Quantum Electrodynamics (circuit-QED), we investigate the DCE for a massless scalar field in a two-dimensional space-time with a time-dependent Robin BC, considering a thermal state as the initial field state. 

The paper is organized as follows: In Sec. \ref{experiments} we describe some experimental proposals related to the DCE. Interesting properties of the Robin BC are discussed in Sec.  \ref{cc-de-Robin}.  The Sec. \ref{modelando-espelho-movel-no-parado} is devoted to the model of a massless scalar field submitted to a time-dependent Robin BC at a static mirror. Final remarks and perspectives are presented in Sec. \ref{conclusao}.

\section{Experimental proposals}
\label{experiments}

The first experimental observation of the DCE was done recently by Wilson {\it et al} \cite{Wilson-Johansson-et-al-arXiv-1105.4720-2011}
in the context of circuit-QED (see Ref. \cite{Nation-J-R-Johansson-Blencowe-F-Nori-arXiv-1003-0835v1-2011} for a review).
This experimental measurement was based on a paper presented by J. R. Johansson {\it et al} \cite{J-R-Johansson-G-Johansson-C-Wilson-F-Nori-PRL-2009} in $2009$, which predicts the generation of photons in a superconducting coplanar waveguide terminated by a SQUID (Super Conducting Quantum Interference Device).
According to \cite{J-R-Johansson-G-Johansson-C-Wilson-F-Nori-PRL-2009},
a time-dependent magnetic flux can be applied on the system, changing the effective inductance of the SQUID, resulting in a time-variable boundary condition, such that the coplanar waveguide becomes equivalent to a transmission line with variable length.
This setup simulates a single moving mirror whose effective velocity can achieve 10{\%} of speed of light.

Still in the context of circuit-QED, other theoretical and experimental works discussed the DCE in electrical circuits based on a microwave cavity coupled with microfabricated waveguides \cite{J-R-Johansson-G-Johansson-C-Wilson-F-Nori-PRA-2010,Wilson-et-al-PRL-2010}. In the experiment reported in \cite{Wilson-et-al-PRL-2010} it was not possible to distinguish whether the photon generation was initiated by quantum or classical fluctuations. However, in the experimental apparatus used by Wilson {\it et al} \cite{Wilson-Johansson-et-al-arXiv-1105.4720-2011}, the contribution of the thermal photons was extremely small since the system was cooled to a temperature of the order of $50 \; \text{mK}$. The output photon flux was measured at two different temperatures ($50 \; \text{mK}$ and $250 \; \text{mK}$) and the authors were able to conclude that the observed signal was indeed dominated by vacuum and not thermal fluctuations \cite{Dalvit-Nature-News-2011}.

It is important to emphasize that other experimental proposals have been made for the detection of the DCE.
The so called MIR experiment, proposed by Braggio {\it et al} \cite{Braggio-EPL-2005,Agnesi-Braggio-et-al-JPA-2008,Agnesi-JPCS-2009}, consists in a superconducting cavity where one of the walls is covered by a semi-conducting slab illuminated by ultrafast laser pulses. This system simulates a moving mirror, since the semiconductor slab switches from a completely transparent medium to a completely reflective one
when illuminated by an appropriate train of laser pulses.
%

Kim {\it et al} \cite{Kim-Brownell-Onofrio-PRL-2006} proposed an experimental setup for generation and detection of photons by means of the mechanical motion of a film bulk acoustic resonator. The film can vibrate at high frequencies, of order of $3.0 \; \text{GHz}$, producing the dynamical Casimir photons. The generated photons interact with an excited population of ultracold alkali-metal atoms and the photodetection occurs as a superradiance phenomenum in the radio-frequency range. 
%

Another experimental setup involving no mechanical motion was proposed by Dezael and Lambrecht \cite{Dezael-Lambrecht-EPL-2010}. A Casimir-like radiation is predicted due to the interactions of a type-I optical parametric oscillator with a thin non-linear crystal slab inside. This system results in an apparent motion of the mirrors, and the Casimir signal is appreciably higher than that for the case of a mechanical motion.

Kawakubo and Yamamoto \cite{Kawakubo-Yamamoto-PRA-2011} investigated the DCE in a resonant cavity and proposed an experimental setup for its detection by means of a non-stationary plasma mirror with Rydberg atoms. The photon creation could be detected by an excitation process of Rydberg atoms through the atom-field interaction.

Finally, Faccio and Carusotto \cite{Faccio-Carusotto-EPL-2011} presented a recent experimental proposal, in which an appropriate train of laser pulses applied perpendicularly to a cavity, made of non-linear optical fiber, modulates in time the refractive index of the medium filling the cavity. As a consequence, it is expected the observation of Casimir photons in the near-infrared domain. 

Additional information on experimental proposals for the DCE can be found in Ref. \cite{V-V-Dodonov-Phys-Scrip-2010}.

\section{Robin boundary conditions}
\label{cc-de-Robin}

In this section, we briefly discuss some properties of Robin BC. For a scalar field $\phi$, the Robin BC is defined as
\begin{equation}
\Big( \phi - \gamma_{0}\frac{\partial\phi}{\partial n} \Big)\Big|_{\text{boundary}} = 0\;,
\label{Robin-BC}
\end{equation}
where $\gamma_0$ is the time-independent Robin parameter (or simply Robin parameter), a constant with dimension of length. 
Robin BC have interesting properties (see \cite{Mintz-Farina-Maia-Neto-Robson-JPA-2003-I} and references therein). The parameter $\gamma_0$ allows a continuous interpolation between Dirichlet ($\gamma_0 \rightarrow 0$) and Neumann BC ($\gamma_0 \rightarrow \infty$).
Robin BC can simulate the plasma model in real metals for low frequencies and was used long ago as a phenomenological model for penetrating surfaces \cite{Mostepanenko-Trunov-1985}. For $\omega\ll\omega_P$, the parameter $\gamma_0$ plays the role of the plasma wavelength which is directly related to the penetration depth of the field (for a simple demonstration of this result see the Appendix of Ref. \cite{Silva-Farina-arXiv-1102-2238v1-2011}). A detailed discussion of the Casimir effect involving Robin BC can be found in \cite{Romeo-Saharian-2002}.  Further investigations involving Robin BC in the context of the static Casimir effect (for instance, thermal corrections, Casimir piston setups, mathematical developments on possible consistent BC) have appeared recently in the literature \cite{FurtherInvestigations-Robin-BC}

In the context of the DCE, Robin BC appeared for the first in the papers by Mintz and collaborators 
\cite{Mintz-Farina-Maia-Neto-Robson-JPA-2003-I,Mintz-Farina-Maia-Neto-Robson-JPA-2003-II}, where a massless scalar field 
in 1+1 dimensions in the presence of one moving mirror was considered. The authors have shown that, for Robin BC, the dynamical Casimir force acquires a dispersive part (the susceptibility $\chi(\omega)$  acquires a real part). Besides, they have shown that the particle creation rate can be significantly reduced when compared to the Dirichlet and Neumann cases if one chooses appropriately the values of the Robin parameter and the  frequency of the mechanical  moving plate.

A time-dependent Robin BC is obtained when the Robin parameter is considered as a function of time, namely,
\begin{equation}
\Big(\phi - \gamma\left(t\right) \frac{\partial\phi}{\partial n} \Big)\Big|_{\text{boundary}} = 0\;,
\label{time-dependent-Robin-BC}
\end{equation}
where $\gamma\left(t\right)$ is the time-dependent Robin parameter.
The DCE for a static mirror with a time-dependent Robin BC was recently investigated by Silva and Farina
\cite{Silva-Farina-arXiv-1102-2238v1-2011}. In the following section, we discuss some consequences of the time-dependent Robin BC in the DCE.

\section{Modeling a moving mirror by a static one}
\label{modelando-espelho-movel-no-parado}

This section is devoted to a simple theoretical model that describes static surfaces with time-dependent material properties. Since the Robin parameter is related to the penetration depth of the field, a natural model for simulating moving surfaces is to consider time-dependent Robin parameter.

We start by considering a massless scalar field in $1+1$ dimensions submitted to a Robin BC with a time-dependent Robin parameter at $x=0$:
\begin{equation}
\phi(0,t) \; =\; \left.\gamma(t)\frac{\partial\phi(x,t)}{\partial x}\right|_{x=0}\;.
\label{time-dependent-Robin-BC-fixed-at-origin}
\end{equation}
In order to apply the Ford-Vilenkin perturbative approach \cite{Ford-Vilenkin-PRD-1982}, we assume
\begin{equation}
\gamma\left(t\right)=\gamma_0+\delta\gamma\left(t\right)\, ,\;\;\;\;\vert\gamma(t)\vert\ll\gamma_0  \;\;\;(\gamma_0>0)\, ,
\label{Una-perturbacion-ao-parametro-del-Robin}
\end{equation}
where $\delta\gamma\left(t\right)$ is a prescribed function of $t$ that vanishes in the remote past and distant future. After a straightforward calculation \cite{Silva-Farina-arXiv-1102-2238v1-2011}, we obtain the Bogoliubov transformation between the input and output creation and annihilation operators,
\begin{eqnarray}
a_{\text{out}}(\omega)&=&
a_{\text{in}}(\omega)-2i\sqrt{\frac{\omega}{1+\gamma_0^2\omega^2}}
\int_{-\infty}^{+\infty}\,\frac{d\omega^{\prime}}{2\pi}
\sqrt{\frac{\omega^{\prime}}{1+\gamma_0^2{\omega^{\prime}}^2}}\times \cr\cr
&\times &
\Bigl[\Theta(\omega^{\prime})a_{\text{in}}(\omega^{\prime})
-\Theta(-\omega^{\prime})
a^{\dagger}_{\text{in}}(-\omega^{\prime})
\Bigr] \,
\delta\Gamma(\omega-\omega^{\prime}),
\label{input-output-operators}
\end{eqnarray}
where $\delta\Gamma(\omega)$ is the Fourier transformation of $\delta\gamma(t)$.
Observe the presence of the creation operator $a_\text{in}^\dag(-\omega^\prime)$ in the right side of Eq. (\ref{input-output-operators}). It will give a non null contribution for the average $\langle {a_\text{out}^\dag \, a_\text{out}} \rangle $ of the output operators taken in the initial field state. This average is directly related to the spectral distribution of the created particles with frequency between $\omega$ and $\omega+d\omega$, namely

\begin{equation}
\frac{dN(\omega)}{d\omega} = \frac{1}{2\pi}
\langle{a^{\dag}_\text{out}(\omega)a_\text{out}(\omega)\rangle}.
\label{spectral-distribution-definition}
\end{equation}

Assuming a thermal bath as the initial field state, we must consider
$
\langle{a^{\dag}(\omega^\prime)a(\omega)\rangle} =
\bar{n}(\omega) \delta (\omega - \omega^{\prime}),
$
where
$
\bar{n}(\omega) = 1/(e^{\omega/T}-1)
$
and $T$ being the absolute temperature. For this reason, it is not difficult to show that the spectral distribution is composed by 

\begin{equation}
\frac{dN(\omega)}{d\omega} =
\frac{dN_{{\,}\!\!_{Planck}}(\omega)}{d\omega} + \frac{d\tilde{N}(\omega)}{d\omega},
\label{dN-domega-total}
\end{equation}
where
\begin{equation}
 \frac{d\tilde{N}(\omega)}{d\omega} = \frac{dN_{vac}(\omega)}{d\omega} + \frac{dN_{T}(\omega)}{d\omega} .
\label{dN-domega-total-II}
\end{equation}
The first term in the right hand side (rhs) of (\ref{dN-domega-total}) is due to the density of particles already present in the thermal bath (Planck term).
The first term of the rhs of (\ref{dN-domega-total-II}) represents the density of particles created in the vacuum state, whereas the second term corresponds to the additional particles created due to the presence of the thermal bath.

\begin{figure}[!h]
\centering
\includegraphics[scale=0.7]{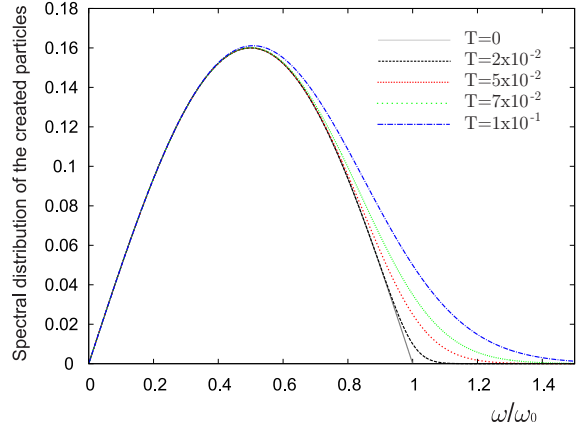}
\caption{\footnotesize (Color online) Spectral  distribution of the created  particles  ${\;}$ $[(2\pi/\epsilon_0^2\tau)d\tilde{N}(\omega)/d\omega]$ ${\;}$ as a function of $\omega/\omega_0$ for $\gamma_0 = 1$ and different values of $\text{T}$ (in arbitrary units). The solid line corresponds to $\text{T} = 0$, the long-dashed line corresponds to $\text{T} = 2 \times 10^{-2}$, the short-dashed line corresponds to $\text{T} = 5 \times 10^{-2}$, the dotted line corresponds to $\text{T} = 7 \times 10^{-2}$ and the dash-dotted line corresponds to $\text{T} = 1 \times 10^{-1}$.}
\label{non-relativistic-spectral-density-thermal-corrections-gamma-1}
\end{figure}

\begin{figure}[!h]
\centering
\includegraphics[scale=0.7]{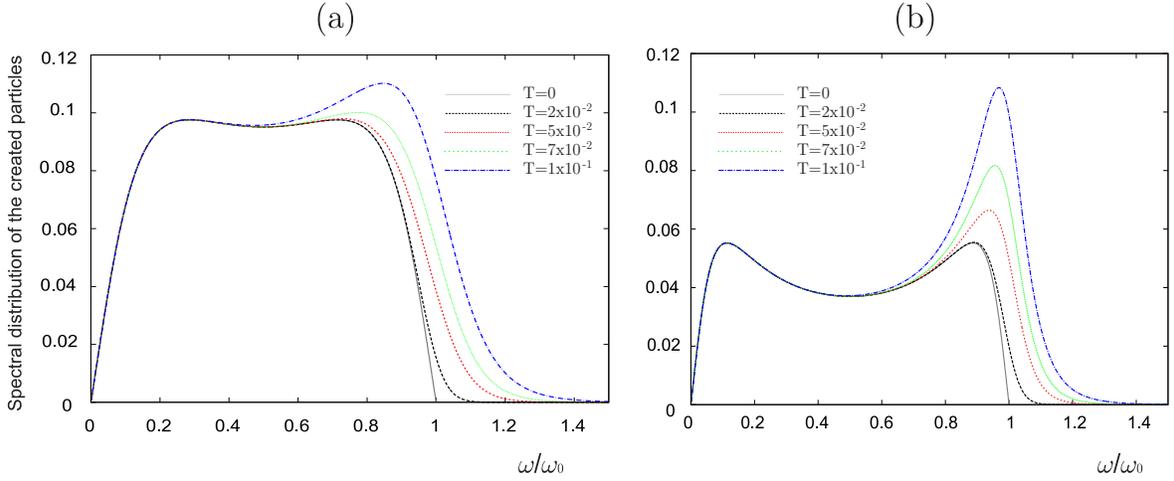}
\caption{\footnotesize (Color online) Spectral distribution of the created particles as function of $\omega/\omega_0$ and different values of $\text{T}$ (in arbitrary units). In $\text{(a)}$, we plot $20\times[(2\pi/\epsilon_0^2\tau)d\tilde{N}(\omega)/d\omega]$ for $\gamma_0 = 5$ and in $\text{(b)}$ $100\times[(2\pi/\epsilon_0^2\tau)d\tilde{N}(\omega)/d\omega]$ for $\gamma_0 = 10$. The solid line corresponds to $\text{T} = 0$, while the long-dashed line, to $\text{T} = 2 \times 10^{-2}$, the short-dashed line,  to $\text{T} = 5 \times 10^{-2}$, the dotted line, to $\text{T} = 7 \times 10^{-2}$ and the dash-dotted line, to $\text{T} = 1 \times 10^{-1}$.}
\label{non-relativistic-spectral-density-thermal-corrections-gamma-5-10-together}
\end{figure}

Inserting (\ref{input-output-operators}) and its hermitian conjugated in Eq. (\ref{spectral-distribution-definition}), we obtain
\begin{equation}
\frac{dN_{vac}(\omega)}{d\omega} = \frac{2}{\pi}
\left( \frac{\omega}{1+\gamma_0^2\omega^2} \right)
\int_{-\infty}^{\infty}\frac{d\omega^{\prime}}{2\pi}\frac{\omega^{\prime}}{1+\gamma_0^2{\omega^{\prime}}^2}{\left\vert
\delta\Gamma(\omega-\omega^{\prime})  \right\vert}^2
\Theta(\omega^{\prime}),
\label{spectral-distribution-general-form}
\end{equation}
\begin{equation}
\frac{dN_T(\omega)}{d\omega} = \frac{2}{\pi}
\left( \frac{\omega}{1+\gamma_0^2\omega^2} \right)
\int_{-\infty}^{\infty}\frac{d\omega^{\prime}}{2\pi}
\frac{\omega^{\prime}}{1+\gamma_0^2{\omega^{\prime}}^2}
\bar{n}(\omega^{\prime})
{\left\vert\delta\Gamma(\omega-\omega^{\prime})  \right\vert}^2 .
\label{thermal-spectral-distribution-general-form}
\end{equation}
Consider for the time-dependent part of the Robin parameter a typical oscillatory behavior with dominant frequency $\omega_0$, namely,
\begin{equation}
\delta\gamma\left(t\right) = \epsilon_0\cos\left(\omega_0 t\right)e^{-\vert{t}\vert/\tau},
\;\;\;\;\;\omega_0 \tau \gg 1,
\label{time-dependent-Robin-parameter-perturbation}
\end{equation}
which substituted into the previous equations leads to the following expressions:
\begin{equation}
 \frac{dN_{vac}(\omega)}{d\omega}
 = \left(\frac{\epsilon^{2}_{0} \tau}{2\pi}\right)
  \frac{\omega\,(\omega_{0}-\omega)}{(1+\gamma_{0}^2\omega^{2})\left[
1+\gamma_{0}^2(\omega_{0}-\omega)^2
\right]}\Theta(\omega_{0}-\omega),
\label{Silva-Farina-vacuum-spectral-distribution}
\end{equation}
\begin{equation}
\frac{dN_T(\omega)}{d\omega} =
\left(\frac{\epsilon^{2}_{0} \tau}{2\pi}\right)
\frac{\omega\,(\omega_{0}-\omega)}{(1+\gamma_{0}^2\omega^{2})
\left[1+\gamma_{0}^2(\omega_{0}-\omega)^2\right]}
\bar{n}(\omega_{0}-\omega).
\label{Silva-Farina-thermal-spectral-distribution}
\end{equation}
Last formulas give us the spectral distributions of the created particles by a fixed mirror submitted to a time-dependent Robin BC. Eq. (\ref{Silva-Farina-vacuum-spectral-distribution}) was computed by Silva and Farina \cite{Silva-Farina-arXiv-1102-2238v1-2011} and corresponds to the particles created by vacuum fluctuations, whereas Eq. (\ref{Silva-Farina-thermal-spectral-distribution}) represents the thermal corrections (see Figs.~\ref{non-relativistic-spectral-density-thermal-corrections-gamma-1} and~\ref{non-relativistic-spectral-density-thermal-corrections-gamma-5-10-together}).

The particle creation rate can be computed by
\begin{equation}
R =  \frac{1}{\tau}
\int_0^\infty \frac{d\tilde{N}(\omega)}{d\omega}\, d\omega.
\label{p-c-rate}
\end{equation}
For the vacuum contribution, we obtain 
\begin{equation}
R_\text{vac} =
\left(\frac{\epsilon^2_{0}\omega^{3}_{0}}{2\pi}\right)
\frac{(2+\xi^2)\ln(1+\xi^2)-2\xi\arctan(\xi)}{\xi^{4} (4+\xi^2)},
\label{vacuum-p-c-rate}
\end{equation}
with $\xi = \omega_0 \gamma_0$, in agreement with \cite{Silva-Farina-arXiv-1102-2238v1-2011}. We did not find an exact analytical expression for the thermal corrections to the particle creation rate, namely
\begin{equation}
R_T =  \frac{1}{\tau}
\int_0^\infty \frac{dN_T(\omega)}{d\omega}\, d\omega,
\label{T-p-c-rate}
\end{equation}
but it can be numerically computed. 

\begin{figure}[!h]
\centering
\includegraphics[scale=0.8]{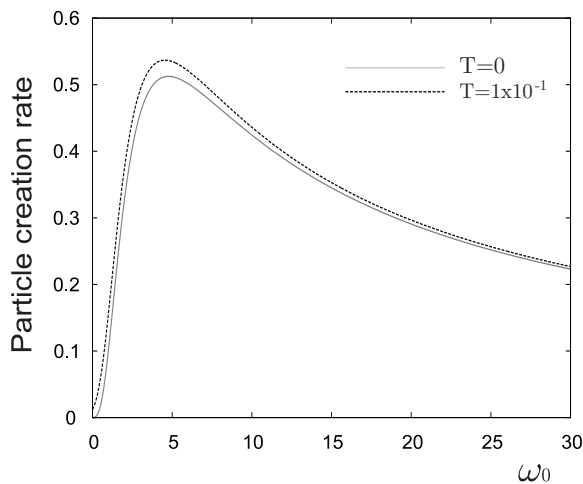}
\caption{\footnotesize (Color online) Particle creation rate $(2\pi/\epsilon_0^2) R $ as a function of $\omega_0$ for $\gamma_0 = 1$ (in arbitrary units of T), plotted in the range $0 < \omega_0 < 30$. The solid line corresponds to $\text{T} = 0$, while the dashed line, to $\text{T} = 1 \times 10^{-1}$.}
\label{particle-creation-rate-thermal-corrections}
\end{figure}

In Fig.(\ref{particle-creation-rate-thermal-corrections}) we plot the particle creation rate $R$ as function of $\omega_0$. 
We observe increasing values of $R$ as $T$ is enhanced. We also remark that, for $T \neq 0$,  there is a residual particle creation rate.
This occurs because, for $\omega_0=0$, it remains a time variation of the Robin parameter: 
$\delta\gamma(t)=\epsilon_0 e^{-|t|/\tau}$.

\section{Final remarks and perspectives}
\label{conclusao}

Considering a time-dependent Robin boundary condition,
we investigated the particle creation in the presence
of a thermal bath, obtaining an enhancement of the number of created particles. 
In the limit of zero temperature, we recover the result for the vacuum state found in the literature \cite{Silva-Farina-arXiv-1102-2238v1-2011}.
For the vacuum case, the particles are created with maximum frequency $\omega_0$ 
(the oscillation frequency of the Robin parameter), and the spectrum is symmetric
under the change $\omega \rightarrow \omega_0-\omega$. 
In contrast, for the thermal case, we observed that
particles can be created with frequencies larger than $\omega_0$. 
In addition, the symmetry under $\omega \rightarrow \omega_0-\omega$ is broken,
since the thermal correction to the vacuum particle creation 
is more accentuated for frequencies $\omega>\omega_0/2$, and
has small effect for $\omega<\omega_0/2$. This asymmetry is enhanced
as we increase the value of $\gamma_0$, leading to a peak for the distribution
of the created particles around $\omega=\omega_0$.

A perspective which is in course, is to generalize the time-dependent Robin BC considered here by including one more term involving a  time derivative of second order. This generalized BC appeared naturally in the work of  Johansson and collaborators \cite{J-R-Johansson-G-Johansson-C-Wilson-F-Nori-PRL-2009}, though there this term could be neglected since they were considering ``dynamics much more slower than the plasma frequency of the SQUID''. We have preliminary results that suggest that the addition of such a term may give an asymmetric contribution to the spectral distribution of created particles. Maybe, in future experiments, the inclusion of such a term can be of some importance (the asymmetry of the spectral distribution of created particles could be used as an additional signature of dynamical Casimir photons). It is worth mentioning that the use of BC including second order derivatives may request some caution, since the inclusion of such terms could lead to mathematical inconsistencies \cite{Castaneda-Asorey}.

\section*{Acknowledgment}
\label{ack}

The authors are indebted to P.A. Maia Neto and B. Mintz for helpful discussions. C.F. acknowledges M. Asorey and J.M. Mu\~noz-Casta\~neda for enlightening conversations. The authors also thank the brazilian agencies CAPES and CNPq  for partial financial support.


\end{document}